\documentclass{elsart}

\usepackage{amssymb}

\usepackage{epsfig}
\usepackage{array}

\def\lapproxeq{\lower .7ex\hbox{$\;\stackrel{\textstyle 
<}{\sim}\;$}} 
\def\gapproxeq{\lower .7ex\hbox{$\;\stackrel{\textstyle 
>}{\sim}\;$}}

\def\gsim{\mathrel{\rlap{\raise 1.5pt \hbox{$>$}}\lower 3.5pt
\hbox{$\sim$}}}
\def\lsim{\mathrel{\rlap{\raise 2.5pt \hbox{$<$}}\lower 2.5pt
\hbox{$\sim$}}}

\begin{document}

\begin{frontmatter}
\title{Potential of Photon Collider in resolving SM-like scenarios}
\author[Novosibirsk]{I. F. Ginzburg\thanksref{grig}}
\author[Warsaw]{M. Krawczyk\thanksref{Someone}}
\author[Bergen]{P. Osland\thanksref{grn}}

\address[Novosibirsk]{Sobolev Institute of Mathematics, SB RAS,
     Prosp.\ ac.\ Koptyug, 4, 630090 Novosibirsk, Russia}
\address[Warsaw]{Institute of Theoretical Physics, Warsaw University,  
00-681 Warsaw, Poland}
\address[Bergen]{Department of Physics, University of Bergen,
     Allegt.\ 55, N-5007 Bergen, Norway}

\thanks[grig]{Supported by RFBR grants 99-02-17211 and 00-15-96691.}
\thanks[Someone]{Partially supported by KBN Grants No 2P03B01414,
 No 2P03B05119 and the DESY Directoriat.}
\thanks[grn]{Supported by the Research Council of Norway.}

\begin{abstract}
After operations at the LHC and $e^+e^-$ Linear Colliders it may be found 
that a Standard-Model-like scenario is realized. 
In this scenario no new particle will be
discovered, except a single Higgs boson having partial widths or
coupling constants squared with fundamental particles close,
within anticipated experimental uncertainty, to those of the SM. 
Experiments at a Photon Collider can resolve whether the SM
model or e.g.\ the Two Higgs Doublet Model is realized in Nature.

For the SM-like version  of the 2HDM~(II) we study the loop couplings of 
the Higgs boson with $\gamma\gamma$ and $Z \gamma$, and also with gluons. 
The deviation of the two-photon width from its SM value 
is generally higher than the expected inaccuracy 
in the measurement of $\Gamma_{\gamma \gamma}$ at a Photon Collider. 
The result is sensitive to the parameters of the Higgs self interaction.
\end{abstract}

\end{frontmatter}
\section{Introduction}
It could happen that no new particles will be discovered
the the Tevatron, the LHC and $e^+e^-$ Linear Colliders \cite{Accomando}
except the SM-like Higgs boson. 
In this case the main task for new colliders 
will be to search for signals of new physics via deviations of
observed quantities from Standard-Model predictions.  The study
of Higgs boson production at a Photon Collider \cite{GKST}
offers excellent
opportunities for this \cite{sitges}. 
Indeed, in the SM and in its extensions, all
fundamental charged particles contribute to the $h\gamma \gamma$
and $hZ\gamma$ effective couplings.  Besides, these couplings
are absent in the SM at tree level, appearing only at the loop
level. Therefore, the background for signals of new physics will
be relatively lower here than in processes which are
allowed at tree level of the SM.

In this analysis \cite{gko} we assume that an SM-like scenario is realized,
i.e., the Higgs particle has been found at the Tevatron or the
LHC and its partial widths or coupling constants squared are
precisely measured (mainly at the $e^+e^-$ Linear Collider),
being close to those of the SM within the anticipated
experimental accuracies. This can happen not only in the SM, but also
if Nature is described by some other theory, for example,
the Two Higgs Doublet Model (2HDM) or the Minimal Supersymmetric
Standard Model (MSSM). In the latter cases the observed
Higgs boson could be either one of the two neutral scalars.
Here  we compare the SM and the SM-like scenario in the 2HDM~(II).
\section{Standard-Model-like scenario}
The {\it SM-like scenario} can be defined by the following criteria:
\begin{itemize}
\item
One Higgs boson will be discovered with mass above today's limit 
for an SM Higgs boson \cite{LEPC}, $M_h>113$~GeV. 
This can be either the Higgs boson of the SM or 
one Higgs boson from 
the two neutral CP-even scalars $h$ and $H$ ($M_h < M_H$) of the 2HDM 
or the MSSM.
\item The measured decay widths of this Higgs boson (or
coupling constants squared) to quarks, charged leptons, 
EW gauge bosons and gluons, $\Gamma_i^{\rm exp}$ ($i=q,l,W,Z,g$), 
will be in agreement with their SM values $\Gamma_i^{\rm SM}$ 
within the experimental precision
\begin{equation}
\left|\frac{\Gamma_i^{\rm exp}}{\Gamma_i^{\rm SM}}-1\right|
\ll 1.
\label{widthest}
\end{equation}
\item  No other Higgs boson will be discovered.
Any other Higgs boson is weakly coupled with the $Z$ boson, 
gluons and quarks, or sufficiently heavy:
\begin{equation}
M_H,\quad M_A,\quad
M_{H^{\pm}}> {}(800\mbox{ GeV})  \label{higgslim}
\end{equation}
to escape observation \cite{lhc}.
\item
Any other new particle that may exist is beyond
the discovery limits of LHC and the $e^+e^-$ Linear Collider.
\end{itemize}
\section{Anticipated precision of measured Higgs couplings
in the SM-like scenario}
Let $\delta_i$ be the relative experimental
uncertainty in the partial width, $\Gamma_i^{\rm exp}$ (or
coupling constants squared)
\begin{equation}
\delta_i=\frac{\delta\Gamma_i^{\rm exp}}{\Gamma_i^{\rm exp}}.
\end{equation}

At the TESLA $e^+e^-$ collider the
discussed production cross sections are expected to be measured
with a significantly higher precision than at the LHC \cite{lhc}.
At $M_h\le 140$ GeV and with integrated luminosity 500--1000~fb$^{-1}$
one can expect \cite{Accomando}:
\begin{equation}\begin{array}{l}
\delta_b=0.027, \qquad \delta_\tau=0.062,
\qquad  \delta_c=0.137, \qquad \delta_t=0.055,\\
\delta_Z=0.01, \qquad  \delta_W=0.054,
\qquad  \delta_g=0.06, \qquad \delta_\gamma=0.14.
\end{array}\label{couplacc}
\end{equation}
Experiments at Photon Colliders open new perspectives. In
particular, even with a modest integrated luminosity of
a $\gamma\gamma$ collider in the high energy peak
of about $40$~fb$^{-1}$, a $\gamma\gamma$ collider
makes it possible to improve on the accuracy in measuring the
$h\gamma\gamma\,$ width up to \cite{JikS}:
$$
\delta_\gamma= 0.02\quad \mbox{ for }M_h<140\mbox{ GeV}.$$
The accuracy in the measurement of the effective $hZ\gamma$
($HZ\gamma$) coupling in the process $e\gamma\to eh$
($e\gamma\to eH$) is evidently not so high.

We will use the above uncertainties to constrain  ratios of
actual (in principle measurable) coupling constants of each neutral
Higgs scalar $\phi$ ($h$ or $H$)\footnote{Discussing both these scalars, 
we use the notation $\phi$ for $h$ and $H$.}
with particle $i$ 
to the corresponding value for the Higgs boson in the SM,
\begin{equation}
\label{Eq:chi-def}
\chi_i^\phi= \frac{g_i^\phi}{g_i^{\rm SM}}\,.
\end{equation}

In the SM-like scenario, for the observed Higgs boson $\phi$ all
$|\chi_i|$ are close to 1,
\begin{equation}
\chi_i^{\rm obs}=\pm (1-\epsilon_i),\quad
\mbox{with }  |\epsilon_i|\ll 1\,.
\label{estacc}
\end{equation}
The allowed ranges for $\epsilon_i$ are constrained by the
experimental accuracies $\delta_i$, 
$|\epsilon_i|\le \delta_i$ but there are also additional constraints 
to these $\epsilon_i$ which follow from the structure of the considered 
model.
\section{Two-Higgs-Doublet Model (II)}
We here consider the CP-conserving 
Two-Higgs-Doublet Model in its Model II implementation, denoted by
2HDM~(II) \cite{Hunter,barroso,Haber}. Here, one doublet of
fundamental scalar fields couples to $u$-type quarks, the
other to $d$-type quarks and charged leptons. The Higgs sector
contains three neutral Higgs particles, two CP-even scalars $h$ and
$H$, and one CP-odd (pseudoscalar) $A$, and charged Higgs bosons
$H^\pm$, it coincides in the 2HDM~(II) and in the MSSM.

In the SM-like scenario realized in the 2HDM we need to consider
both possibilities: not only the light scalar Higgs boson, $h$,
but also the heavier one, $H$, could imitate the SM Higgs boson if the
lighter scalar $h$ escapes detection \cite{light-Higgs}, see also \cite{GUN}.

The ratios of the direct coupling constants of the Higgs boson
$\phi=h$ to the gauge bosons $V=W$ or $Z$ bosons, to up and 
down quarks and to charged leptons, relative to their SM values 
can be expressed via angles $\alpha$ and $\beta$ \cite{Haber,Hunter}:
\begin{equation}
\begin{array}{l}
\chi_V^h=\sin(\beta-\alpha), \\
\chi_u^h=\sin(\beta-\alpha)+\cot\beta\cos(\beta-\alpha), \\
\chi_d^h =\sin(\beta-\alpha)-\tan\beta\cos(\beta-\alpha),
\end{array}
\label{2hdmcoup-h}
\end{equation}
with similar expressions for $\phi=H$.
Here $\beta$ parameterizes the ratio of the vacuum expectation
values of the two basic Higgs doublets and $\alpha$ parameterizes
mixing among the two neutral CP-even Higgs fields. 
The angle $\beta$ is chosen in the range $(0,\,\pi/2)$ and 
the angle $\alpha$ in the range $(-\pi,\;0 )$.

The coupling of the charged Higgs boson to the
neutral scalars $\phi$ depends on the Higgs-boson masses
and on the additional parameter $\lambda_5$ \cite{Djouadi1}. 
\section{Two-Higgs-Doublet Pattern relation}
The quantities $\chi_i^\phi$ for the couplings of each
scalar (\ref{2hdmcoup-h}) (referred to below as basic couplings)
are closely related to the observables and in the forthcoming 
analysis it is more natural to use them, instead of 
$\alpha$ and $\beta$.
Since for each $\phi$ these three $\chi_i$ can be expressed in
terms of {\it two} angles, they fulfill a simple relation
{\em(pattern relation)}, which plays a basic role in our analysis. 
It has the same form for  both $h$ and $H$, namely $(\chi_u -\chi_V)
(\chi_V -\chi_d) +\chi_V^2=1$, or 
\begin{equation}
(\chi_u +\chi_d)\chi_V=1+\chi_u \chi_d.
\label{2hdmrel}
\end{equation}
Furthermore, from Eq.~(\ref{2hdmcoup-h}) follows an expression 
for $\tan^2\beta$:
\begin{equation}
\label{Eq:tan-beta}
\tan^2\beta={\frac{\chi_V-\chi_d}{\chi_u-\chi_V}}
={\frac{1-\chi_d^2}{\chi_u^2-1}}.
\end{equation}

In the following discussion we will assume only one value for each
up-type quark, down-type quark, charged lepton and gauge boson coupling
with the Higgs boson, in numerical calculation we will use the best 
estimate for each category, e.g. $\delta_b$ for $\delta_d$, etc.
\section{Allowed ranges for couplings}
The SM-like scenario means, in particular, that 
$\chi_i^2\approx 1$ (here, we consider only basic couplings
with $i=u,d,V$; loop couplings are discussed in the next section). 
We consider solutions of the equations (\ref{estacc})
constrained by the pattern relation (\ref{2hdmrel}). Taking into account 
the definitions (\ref{2hdmcoup-h}) and earlier noted regions of variation 
of the mixing angles $\alpha$ and $\beta$, we check if the obtained 
solution can be classified as SM-like with respect to the first criterion 
of Sec.~2.
For the observed Higgs boson $\phi$ we consider solutions
denoted $A_{\phi 1}$ and $A_{\phi 2}$ with approximately 
identical $\chi_V\approx \chi_u\approx \chi_d\approx \pm 1$. 
Subscript 1 corresponds to solutions with $\chi_V\approx 1$ while 
subscript 2 labels solutions with $\chi_V\approx -1$. 
These solutions are really close to the SM for all 
basic couplings: relative phases coincide, and magnitudes are practically
the same.
There are also solutions where some of the $\chi_i\approx 1$ 
but other $\chi_j\approx -1$.
These are denoted $B_{\phi 1}$ and $B_{\phi 2}$, where the 
subscripts have the same meaning as above.
These solutions are in fact distinct from the SM case, even though
all basic widths $\sim\chi_i^2$ are close to the SM values; we will not
discuss them here, for details see \cite{gko}.

For the solutions $A_{Hi}$ where the observed SM-like 
Higgs boson is the heavier one, the values for the basic couplings 
can describe a picture 
which is different from the SM-like scenario since 
the lighter Higgs boson can in principle be observable. We 
discuss these possibilities in each case and exclude 
some solutions from the subsequent discussion, if they do not fulfill 
the criteria of Sec.~2 (no other Higgs particle should be discovered).

For the solutions A near the SM point for all basic
coupling constants: $\chi_V=\chi_u=\chi_d=\pm 1$, i.e.,
\begin{equation}\begin{array}{cl}
\chi_V=1-\epsilon_V, \quad  \chi_d=1-\epsilon_d, \quad
\chi_u=1-\epsilon_u, \qquad  &(A_{h1},\;A_{H1})\\ 
\chi_V=-1+\epsilon_V,\quad  \chi_d=-1+\epsilon_d, \quad 
\chi_u=-1+\epsilon_u. \qquad &(A_{h2},\;A_{H2})\, ,
\end{array}\label{AH}
\end{equation}
we obtain, using the pattern relation and
neglecting terms of higher order in $\epsilon_i$,
\begin{equation}
\tan\beta=\frac{|\epsilon_d|}{\sqrt{2\epsilon_V}}
=\sqrt{\left|\frac{\epsilon_d}{\epsilon_u}\right|}, \quad
\epsilon_u=-\frac{2\epsilon_V}{\epsilon_d}
\Rightarrow \epsilon_V=-\frac{\epsilon_u\epsilon_d}{2}
<\frac{\delta_u\delta_d}{2}.\label{sol2s}
\end{equation}

According to Eqs.~(\ref{estacc}) and (\ref{2hdmcoup-h}), 
$\epsilon_V>0$, so that in
all these solutions the signs of $\epsilon_u$ and $\epsilon_d$ are
opposite. Since $\epsilon_V$ here is given by
the product of two other $\epsilon$'s, it should be extremely
small, ($\epsilon_V\le0.001$ using $\delta_t$ and $\delta_b$ from
Eq.~(\ref{couplacc}), while for $V=Z$ we expect $\delta_Z$=0.01). 
This deviation can therefore be neglected, and in
the calculations of loop-induced couplings 
one can put $\chi_V=\pm 1$.
\section{Distinguishing models via loop couplings}
In order to distinguish models in the considered SM-like scenario, we
compute loop-induced couplings of the Higgs boson with photons or
gluons \cite{Hunter,Djouadi1} for solutions $A_h$ and $A_H$
within the ranges of the coupling constants allowed by 
the anticipated experimental inaccuracies from Eq.~(\ref{couplacc})
and within the constraints of the pattern relation, Eqs.~(\ref{AH})
and (\ref{sol2s}).
To estimate the deviation from the SM, 
we consider the ratios of widths
$\chi^2_{\gamma\gamma}$ and $\chi^2_{Z \gamma}$ 
obtained in the 2HDM~(II) and in the SM. In the 2HDM
the couplings with photons, $h\gamma\gamma$ and $hZ\gamma$, contain
contributions from fermions, and from the charged gauge boson $W^{\pm}$, 
like in the SM. 
In addition, there are contributions from the charged Higgs boson, 
$H^{\pm}$.

For definiteness, we perform all calculations here for
$M_{H^\pm}=800$~GeV. At $M_\phi<250$~GeV the contribution of
the charged Higgs boson loop varies by less than 5\% when $M_{H^{\pm}}$ 
varies from 800~GeV to infinity. 

The $\gamma\gamma$ and $Z\gamma$ widths look the most
promising ones for distinguishing models. A new feature of these
widths, as compared to the SM case, is the contribution due to
the charged Higgs boson loops. It is known that the scalar loop
contribution to the photonic widths is less than that of fermion
and $W$ boson loops (the last is the largest). 
The contributions of $W$ and $t$-quark loops are of opposite sign,
i.e., they partially compensate each other, thus, the effect of
scalar loops is enhanced here.
The coupling $\chi_{H^{\pm}}$ depends on $\lambda_5$, 
which is not fixed by the observable masses. 
This dependence is linear in the considered partial widths 
\begin{equation}
\chi_a^2 =\frac{\Gamma_a^{\rm 2HDM}}{\Gamma_a^{\rm SM}}
=1-R_a(1-\Lambda), \qquad a=\gamma\gamma\mbox{ or } Z\gamma,
\label{Ra}
\end{equation}
with $\Lambda$ proportinal to $\lambda_5$.

For $\lambda_5=0$ the ratios of the
considered Higgs widths to their SM values are shown in
Fig.~1. Here, thick curves correspond to strict SM
values for the basic couplings with quarks and gauge bosons.
These curves are below unity due to the contribution of the charged
Higgs boson. The shaded regions are derived from the anticipated
1~$\sigma$ bounds around the SM values of {\sl two} measured
basic coupling constants, $g_b$ and $g_t$
(the third basic coupling $g_Z$ must be very close to its SM value). 
Solutions A cover the shaded area in Fig.~1.
With increasing Higgs boson mass, the deviation of the considered loop
couplings from their SM values decreases monotonically. 
\begin{figure}[htb]
\refstepcounter{figure}
\label{Fig:sol-A}
\epsfig{file=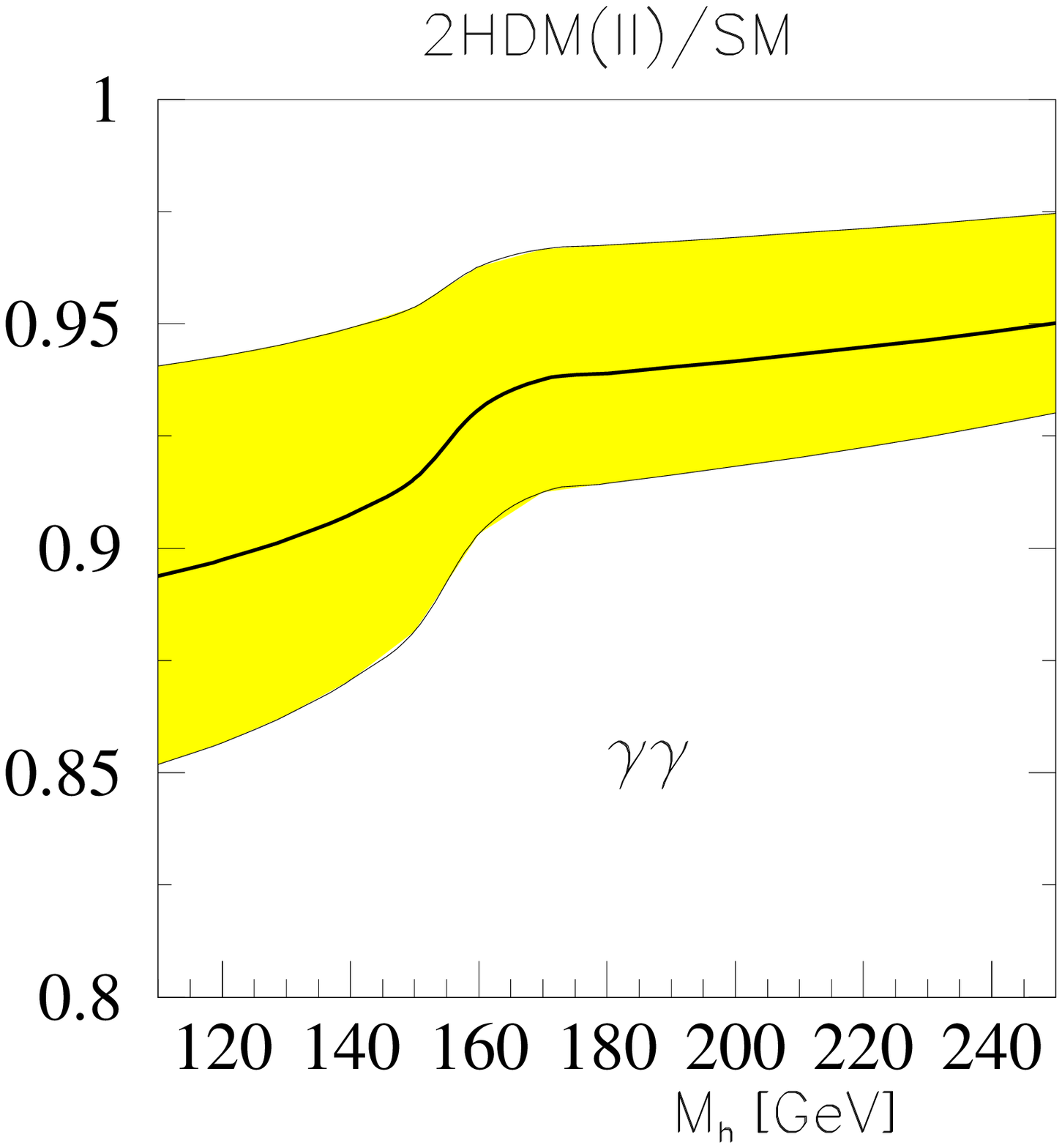,width=72mm}
\epsfig{file=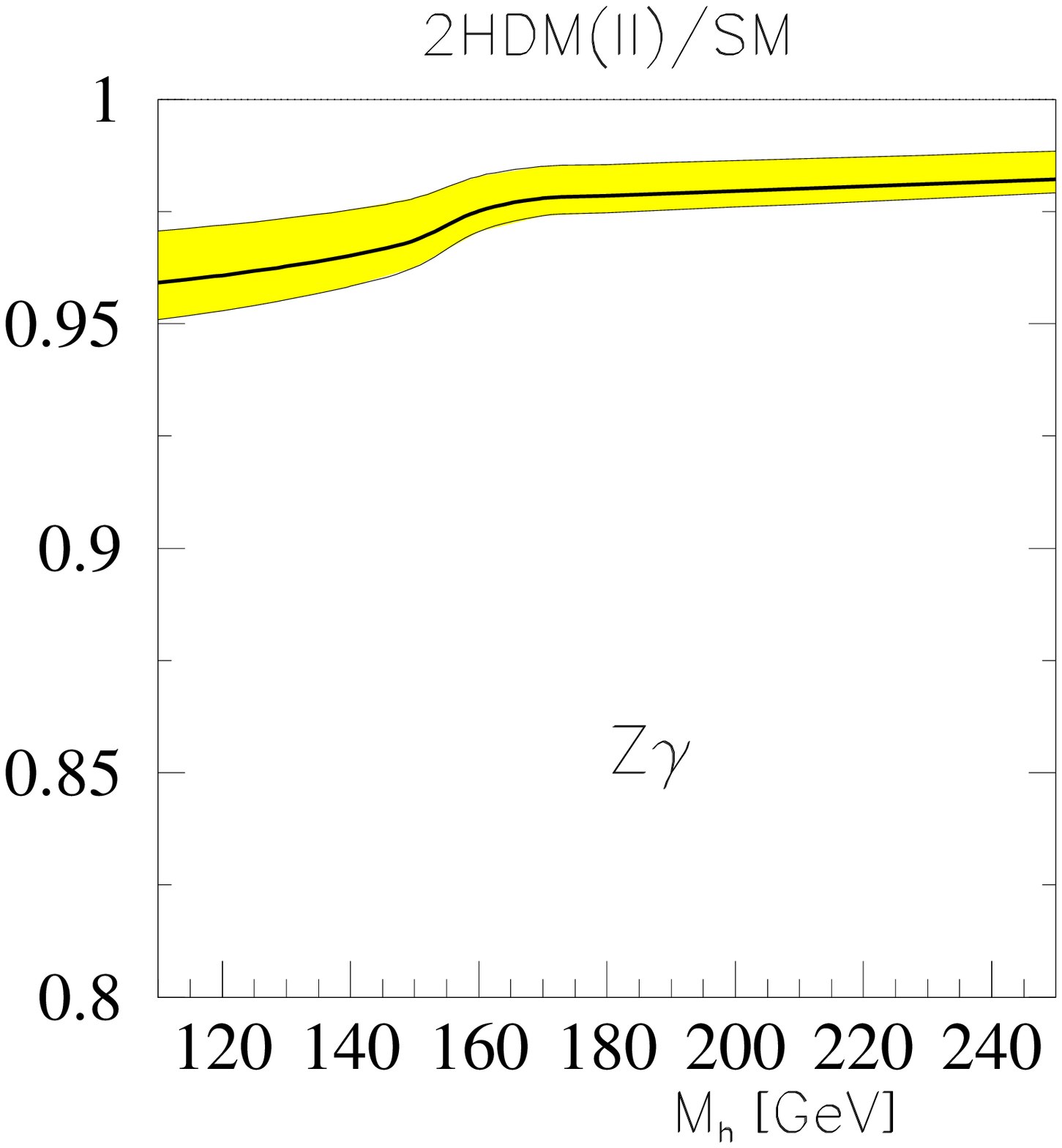,width=72mm}
\addtocounter{figure}{-1}
\caption{Ratios of the Higgs boson $\phi\to\gamma\gamma$ 
and $\phi\to Z\gamma$ decay widths in the
2HDM and the SM as functions of $M_h$ for all solutions~$A$ with
$\lambda_5=0$. See text for description of the shaded bands.}
\end{figure}

We see that the deviation from unity is large enough to allow 
a reliable distinction of the 2HDM from the SM in the process 
$\gamma\gamma\to H$.
This conclusion is valid in a wide range of $\lambda_5$ values.
The possible precision in the determination of $\lambda_5$ from 
the two-photon width depends crucially 
on the mass of the charged Higgs boson.

The two-gluon width is determined by the contributions of
$t$ and $b$ quarks. For not too high values of $\tan\beta$, the
$t$-quark contribution dominates. So, the difference 
$\chi_{gg}-1$ is determined by the difference $\chi_u-1$, and 
with high accuracy $\chi_{gg}-1 \approx 2(\chi_u-1)$.
If $\tan\beta\ll 1$ then the deviation of the Higgs boson 
coupling with $t$-quark from its SM value can be large 
compared to expected experimental uncertainty (\ref{couplacc}). 
In this case the two-gluon width can differ from its SM value 
by more than the experimental uncertainty, and
the measurement of the two-gluon width could exclude the SM-like scenario
from being realized by the 2HDM.
In such a case the Photon Collider can be used for a more detailed study 
of the realized model beyond the SM.
\section{Conclusion}
An SM-like scenario observed  at the LHC and 
$e^+e^-$ Linear Colliders can occur both in the SM
and in other models, including the 2HDM~(II). 
In order to distinguish these models,
we implement a pattern relation among basic couplings.
Taking into account anticipated uncertainties in future measurements
of the basic couplings of
the Higgs boson, we found that the pattern relation in the considered 
SM-like scenario, in which partial widths
of Higgs boson decay are close to their SM values,
has two types of solutions.
In the solutions $A$ all basic couplings are close to their SM values. 
(In the solutions $B$, 
some of the basic couplings are close to their SM values 
while others differ in sign from the SM values.)

We studied the $\gamma\gamma$ and $Z\gamma$ partial widths of the 
observed Higgs boson for all considered solutions. 
The obtained predictions are practically identical for the cases 
when the observed 
scalar is the lighter Higgs boson of 2HDM ($h$) or the heavier one ($H$). 
For solutions $A$ this difference depends on an additional parameter of
the theory, $\lambda_5$. 
These solutions can be discriminated in a considerable range
of values of this parameter. The $\gamma\gamma$ 
width in the 2HDM~(II) differs significantly from the corresponding 
SM width in a wide range of parameters. 
Therefore, with the anticipated high accuracy of 
measuring this width at a Photon Collider, such a measurement could 
in general resolve the 2HDM~(II) and the SM.

\end{document}